\documentclass[12pt]{article}

\usepackage{amsmath,amsfonts}
\usepackage{cite}
\allowdisplaybreaks[1]

\title{BRST approach to Lagrangian construction\\ for bosonic continuous spin field}
\author{
I.L. Buchbinder$^{a,b,c}$\footnote{joseph@tspu.edu.ru},
V.A. Krykhtin$^a$\footnote{krykhtin@tspu.edu.ru },
H. Takata$^a$\footnote{takata@tspu.edu.ru }
\\[1em]
{\it $^a$Department of Theoretical Physics, Tomsk State Pedagogical  University,}\\ {\it  Kievskaya St.\ 60, 634061, Tomsk,  Russia }
\\[1em]
{\em $^b$National Research Tomsk State  University,}\\{\em Lenin Av.\ 36, 634050 Tomsk, Russia}
\\[1em]
{\em    $^c$Departamento de F\'isica, ICE, Universidade Federal de Juiz de Fora,}\\
{\em Campus Universit\'ario-Juiz de Fora, 36036-900, MG, Brazil}
}
\date{}

\begin{document}
\maketitle

\begin{abstract}
We formulate the conditions defining the irreducible continuous spin
representation of the four-dimensional Poincar\' e group based on spin-tensor fields with dotted and undotted indices. Such a
formulation simplifies analysis of the Bargmann-Wigner equations and
reduces the number of equations from four to three. Using this
formulation we develop the BRST approach and derive the covariant
Lagrangian for the continuous spin fields.

\end{abstract}

\section{Introduction}
Description of the irreducible representations of the Poincar\'e and
AdS groups play important role in the formulation of the higher spin
field models (see e.g. the reviews \cite{Vasiliev}, \cite{Didenko}).
One of such representations is representation with continuous spin.

Continuous (or infinite) spin representation of the Poincar\' e
group \cite{Bargmann:1948ck,Iverson-Mack,Brink:2002zx} being
massless has unusual properties such as an infinite number of
degrees of freedom and appearance of the dimensional parameter $\mu$
in the conditions defining the irreducible representation (see for
the review\cite{Bekaert:2017khg}). Lagrangian for the bosonic field
in $d=4$ was first proposed in ref. \cite{Schuster:2014hca} and its
structure was analyzed in ref.  \cite{Rivelles:2014fsa}. Later
Lagrangian for bosonic continuous field was generalized for $d>4$
and written in terms of a tower of double traceless tensor fields in
the works \cite{Metsaev:2016lhs} and in terms of two towers of
traceless fields in the works \cite{Metsaev:2018lth} (see also the
approaches to Lagrangian constriction in the works
\cite{Zinoviev:2017rnj,Khabarov:2017lth,Najafizadeh:2015uxa,Metsaev:2017ytk,
Metsaev:2017,Najafizadeh:2017tin} for bosonic and fermionic fields
and relation to the string theory
in\cite{Savvidy:2003fx,Mourad:2005rt}). Interaction of continuous
spin fields with finite spin massive fields was considered in the
papers \cite{Metsaev:2017cuz,Bekaert:2017xin}. Model of relativistic
particle corresponding to continuous spin field has been constructed
in the recent work \cite{Fedoruk}.

In the present paper we develop the  BRST approach to derive the
Lagrangian for the continuous spin fields in four-dimensional
Minkowski space. This approach is a direct generalization of the
general BRST construction which was used in our works for deriving
the Lagrangians for the free fields of different types in flat and
AdS spaces (see e.g.
\cite{Buchbinder1,Buchbinder:2004gp,Buchbinder:2005ua,Buchbinder:2006nu,Buchbinder:2006ge}
and the references therein, see also the review
\cite{Tsulaia})\footnote{The various applications of the BRST-BFV
construction in the continuous spin field theory have been studied
in refs. \cite{Metsaev:2016lhs,Metsaev:2017ytk,
Metsaev:2017,Metsaev:2017cuz,Metsaev:2018lth,
Bengtsson:2013vra,Alkalaev:2017hvj}.}. The crucial element of our
approach is the implementation of the two-component spinor
description for the continuous spin fields. We suppose that the BRST
construction in terms of spin-tensor fields, considered in the given
paper, essentially simplifies an clarifies the derivation of the
Lagrangian and its analysis.

The paper is organized as follows. In the next section we consider a
new, in comparison with ref. \cite{Bargmann:1948ck}, possibility of
realization for the spin momentum operators in terms of
two-component spinors and obtain new equations for the field
realizing the continuous spin irreducible representation of the
Poincar\'e group. After this we solve one of these equations and in
section~\ref{sec:Lagr} we construct Lagrangian on the base of the
BRST method.  Then we show that after removing one auxiliary field
and rescaling the other fields and the gauge parameter the
Lagrangian obtained exactly coincides with Lagrangian derived by
Metsaev \cite{Metsaev:2018lth}.

\section{Spin-tensor representation}
According to Bargmann and Wigner \cite{Bargmann:1948ck}, the
continuous spin field are characterized by the following
eigenvalues of the Casimir operators
\begin{eqnarray}
P^2\Psi=0 &\qquad& W^2\Psi=\mu^2\Psi.
 \label{CSConditions}
\end{eqnarray}
In order to obtain these equations in explicit form we need the
explicit expressions for the operators entering into the Casimir
operators. Such expressions for the operators depend on how $\Psi$
transforms under action of the Poincar\'e group.

In the works on continuous spin field it usually was done in the
following way. Let us introduce an auxiliary 4-dimensional vector
$w^\mu$ and define
\begin{eqnarray}
\varphi_s(x,w)&=&\varphi_{\mu_1\ldots\mu_s}(x)\;w^{\mu_1}\ldots w^{\mu_s}
,
\label{fxw}
\end{eqnarray}
where $\varphi_{\mu_1\ldots\mu_s}(x)$ is a totally symmetric tensor field.
Then one realizes the spin momentum operator for $\varphi_s(x,w)$ as
follows
\begin{eqnarray}
M_{\mu\nu}=w_\mu i\frac{\partial}{\partial w^\nu}-w_\nu
i\frac{\partial}{\partial w^\mu} .
\label{Susual}
\end{eqnarray}
Using this relation for the spin momentum operator one can write the
Wigner equations \cite{Bargmann:1948ck}
\begin{eqnarray}
&&p^2\Psi(p,w)=0,
\nonumber \\
&&(p_\nu w^\nu+\mu)\Psi(p,w)=0,
\nonumber \\
&&p^\nu\frac{\partial}{\partial w^\nu}\Psi(p,w)=0,
\nonumber \\
&&\left(\frac{\partial}{\partial w^\nu}\frac{\partial}{\partial
w_\nu}+1\right)\Psi(p,w)=0.
\label{Wigner}
\end{eqnarray}
If the four
equations (\ref{Wigner}) are satisfied, then the field $\Psi$ will satisfy
(\ref{CSConditions}) and thus will describe the irreducible representation of the Poincare group with continuous spin.

We want to turn attention that there is another possibility to realize the spin momentum operator and
corresponding representation.
Instead of usual tensor fields (\ref{fxw}) we consider the spin-tensor field
$\varphi_{a_1\ldots a_n\dot{b}_1\ldots\dot{b}_k}(x)$ with $n$ undotted and $k$ dotted indices and define
\begin{eqnarray}
\varphi_{n,k}(x,\xi,\bar{\xi})
&=&
\varphi_{a_1\ldots a_n\dot{b}_1\ldots\dot{b}_k}(x)\;
\xi^{a_1}\ldots\xi^{a_n}\;\bar{\xi}^{\dot{b}_1}\ldots\bar{\xi}^{\dot{b}_k},
\label{spin-tensor}
\end{eqnarray}
where we have introduced two auxiliary bosonic 2-dimensional spinors
$\xi^a$ and $\xi^{\dot{b}}$ of the Lorentz group\footnote{The
two-component spinors have been used for description of the
representations with continuous spin in ref. \cite{Iverson-Mack}.
However they were applied for the other aims.}. In this
representation the spin momentum operator looks like
this\footnote{We use the notation as in book \cite{Ideas}}
\begin{eqnarray}
M_{\mu\nu}&=&\sigma_{\mu\nu}^{ab}M_{ab}-\bar{\sigma}_{\mu\nu}^{\dot{a}\dot{b}}M_{\dot{a}\dot{b}},
\label{M}
\end{eqnarray}
where
\begin{eqnarray}
M_{ab} &=&
\frac{1}{2}\xi^c\varepsilon_{ca}\frac{\partial}{\partial\xi^b}+\frac{1}{2}\xi^c\varepsilon_{cb}
\frac{\partial}{\partial\xi^a} =
-\frac{i}{2}\Bigl(\xi_a\pi_b+\xi_b\pi_a\Bigr),
\label{notations1}
\\
\bar{M}_{\dot{a}\dot{b}} &=&
\frac{1}{2}\bar{\xi}^{\dot{c}}\varepsilon_{\dot{c}\dot{a}}\frac{\partial}{\partial{\bar\xi}^{\dot{b}}}
+
\frac{1}{2}\bar{\xi}^{\dot{c}}\varepsilon_{\dot{c}\dot{b}}\frac{\partial}{\partial{\bar\xi}^{\dot{a}}}
=
-\frac{i}{2}\Bigl(\bar{\xi}_{\dot{a}}\bar{\pi}_{\dot{b}}+\bar{\xi}_{\dot{b}}\bar{\pi}_{\dot{a}}\Bigr)
\label{notations2}
\\
&&
\pi_a=-i\frac{\partial}{\partial\xi^a}
\qquad
\bar{\pi}_{\dot{a}}=-i\frac{\partial}{\partial{\bar\xi}^{\dot{a}}}
\;.
\label{notations3}
\end{eqnarray}
One can prove that the operator (\ref{M}) satisfies the commutation
relation for the Lorentz group generators.

To construct the second Casimir operator for the spin-tensor representation one uses the explicit expression
for the spin operator (\ref{M}) acting on the spin-tensor fields.
In this case the second Casimir operator becomes
\begin{eqnarray}
W^2
&=&
-M_{ab}\,\bar{M}_{\dot{a}\dot{b}}\,P^{a\dot{a}}P^{b\dot{b}}
+\frac{1}{2}\Bigl(M_{ab}M^{ab}+\bar{M}_{\dot{a}\dot{b}}\bar{M}^{\dot{a}\dot{b}}\Bigr)P^2
\end{eqnarray}
After some transformations the second Casimir operator takes the form
\begin{eqnarray}
W^2
&=&
\frac{1}{2}(\xi\sigma^\mu\bar{\xi})(\bar{\pi}\bar{\sigma}^\nu\pi)P_\mu P_\nu
+\frac{1}{2}(\bar{\xi}\bar{\sigma}^\mu\pi)(\bar{\pi}\bar{\sigma}^\nu\xi)P_\mu P_\nu
\nonumber
\\
&&{}
+\frac{1}{2}\Bigl(
M_{ab}M^{ab}+\bar{M}_{\dot{a}\dot{b}}\bar{M}^{\dot{a}\dot{b}}
+i\bar{\xi}^{\dot{a}}\bar{\pi}_{\dot{a}}
\Bigr)P^2
\end{eqnarray}
Using the identity
\begin{eqnarray}
(\bar{\xi}\bar{\sigma}^\mu\pi)(\bar{\pi}\bar{\sigma}^\nu\xi)P_\mu P_\nu
=
(\xi\sigma^\mu\bar{\xi})(\bar{\pi}\bar{\sigma}^\nu\pi)P_\mu P_\nu
+\bar{\xi}^{\dot{a}}\bar{\pi}_{\dot{a}}(i+\pi_a\xi^a)P^2
\end{eqnarray}
we can write the operator $W^2$ in two equivalent forms
\begin{eqnarray}
W^2
&=&
(\xi\sigma^\mu\bar{\xi})(\bar{\pi}\bar{\sigma}^\nu\pi)P_\mu P_\nu
\nonumber
\\
&&{}
+\frac{1}{2}\Bigl(
M_{ab}M^{ab}+\bar{M}_{\dot{a}\dot{b}}\bar{M}^{\dot{a}\dot{b}}
+\bar{\xi}^{\dot{a}}\bar{\pi}_{\dot{a}}\xi^a\pi_a
\Bigr)P^2
\label{W1}
\end{eqnarray}
or
\begin{eqnarray}
W^2
&=&
(\bar{\xi}\bar{\sigma}^\mu\pi)(\bar{\pi}\bar{\sigma}^\nu\xi)P_\mu P_\nu
\nonumber
\\
&&{}
+\frac{1}{2}\Bigl(
M_{ab}M^{ab}+\bar{M}_{\dot{a}\dot{b}}\bar{M}^{\dot{a}\dot{b}}
-\bar{\xi}^{\dot{a}}\bar{\pi}_{\dot{a}}\pi_a\xi^a
\Bigr)P^2.
\label{W2}
\end{eqnarray}

We will consider the irreducible representation with continuous spin on the fields
$\Psi(p,\xi,\bar{\xi})$, depending on the momentum $p_{\mu}$ and spin-tensor variables
$\xi^{a}$ and $\xi^{\dot{a}}.$ Let the field $\Psi(p,\xi,\bar{\xi})$ satisfies the constraints
\begin{eqnarray}
&&
p^2\Psi(p,\xi,\bar{\xi})=0,
\label{C1}
\\
&&
((\bar{\pi}\bar{\sigma}^\nu\pi) p_\nu+i\mu)\Psi(p,\xi,\bar{\xi})=0,
\label{C2}
\\
&&
((\xi\sigma^\mu\bar{\xi})p_\mu-i\mu)\Psi(p,\xi,\bar{\xi})=0.
\label{C3}
\end{eqnarray}
Then on can show that conditions (\ref{CSConditions}) are satisfied and hence the field $\Psi(p,\xi,\bar{\xi})$ describes
the irreducible representation with continuous spin.
Thus we have obtained the equations on the field $\Psi$ in the spin-tensor representation.
Equations (\ref{C1})--(\ref{C3}) are similar to the ``modified Wigner's equations'' in the paper \cite{Najafizadeh:2017tin}.
In the case under consideration analog of the fourth equation in (\ref{Wigner}) are resolved automatically due to the properties
of the two component spinors.

We will construct the Lagrangian for the continuous spin field.
To do that
one should somehow decompose $\Psi(p,\xi,\bar{\xi})$ in a series of $\xi$ and $\bar{\xi}$
and get the spin-tensor fields.
However, one can see that, because of equation (\ref{C3}), $\Psi(p,\xi,\bar{\xi})$ such a direct decomposition
is impossible. Therefore we first solve  (\ref{C3}) in the form
\begin{eqnarray}
\Psi(p,\xi,\bar{\xi})=\delta((\xi\sigma^\mu\bar{\xi})p_\mu-i\mu)\;\varphi(p,\xi,\bar{\xi}).
\end{eqnarray}
One can prove that if the field $\varphi(p,\xi,\bar{\xi})$ obeys
equations
\begin{eqnarray}
&&
\partial^2\varphi(x,\xi,\bar{\xi})=0,
\label{C1+}
\\
&&
\left(\bar{\sigma}^{\mu\dot{a}a}\frac{\partial}{\partial\xi^a}
\frac{\partial}{\partial\bar{\xi}^{\dot{a}}}
\frac{\partial}{\partial  x^\mu}
+\mu\right)\varphi(x,\xi,\bar{\xi})=0
\label{C2+}
,
\end{eqnarray}
then the field $\Psi(p,\xi,\bar{\xi})$ will satisfy the rest equations (\ref{C1}) and (\ref{C2}).
Here we have made Fourier transform from momentum $p^\mu$ representation into the coordinates $x^\mu$ representation.
Equations (\ref{C1+}) and (\ref{C2+}) have a solution in the form of an expansion in $\xi$ and $\bar{\xi}$
\begin{eqnarray}
\varphi(x,\xi,\bar{\xi})
=
\sum_{n,k=0}^\infty
\frac{1}{\sqrt{n!k!}}\,\varphi_{a_1\ldots a_n\dot{b}_1\ldots\dot{b}_k}(x)\;
\xi^{a_1}\ldots\xi^{a_n}\;\bar{\xi}^{\dot{b}_1}\ldots\bar{\xi}^{\dot{b}_k}
.
\label{dec}
\end{eqnarray}
Since we are going to construct Lagrangian for real bosonic fields we will consider $n=k$ case in (\ref{dec}).

\section{Lagrangian construction}\label{sec:Lagr}

Following the general BRST approach in higher spin field theory we
begin with realization of the equations (\ref{C1+}) and (\ref{C2+})
in auxiliary Fock space.

Let us introduce creation and annihilation operators
\begin{align*}
&\langle0|\bar{c}_{\dot{b}}=\langle0|c^a=0,
&&
\bar{a}^{\dot{b}}|0\rangle=a_a|0\rangle=0,
&&\langle0|0\rangle=1
\end{align*}
with the following nonzero commutation relations
\begin{eqnarray*}
[\bar{a}^{\dot{\alpha}},\bar{c}_{\dot{\beta}}]
=\delta^{\dot{\alpha}}_{\dot{\beta}},
&\qquad&
[a_\alpha,c^\beta]=\delta_\alpha^\beta.
\end{eqnarray*}
The states in the auxiliary Fock space are defined as follows
\begin{eqnarray}
|\varphi\rangle=
\sum_{k,l=0}^{\infty}|\varphi_{kl}\rangle
&\qquad&
|\varphi_{kl}\rangle=\frac{1}{\sqrt{k!l!}}\,\varphi_{a(k)}{}^{\dot{b}(l)}(x)\,c^{a(k)}\,\bar{c}_{\dot{b}(l)}|0\rangle.
\label{GFState}
\end{eqnarray}

We determine the Hermitian conjugation in the Fock space by the rule
\begin{align*}
&(a_a)^+=\bar{c}_{\dot{a}}
&&(\bar{c}_{\dot{a}})^+=a_a
&&(\bar{a}_{\dot{a}})^+=c_a
&&(c_a)^+=\bar{a}_{\dot{a}}
\end{align*}
Then the state which is Hermitin conjugate to state (\ref{GFState}) is written as follows
\begin{eqnarray}
\langle\bar{\varphi}|&=&
\sum_{k,l=0}^{\infty}\frac{1}{\sqrt{k!l!}}\,\langle 0|\,\bar{a}^{\dot{a}(k)}\,a_{b(l)} \bar{\varphi}^{b(l)}{}_{\dot{a}(k)}.
\end{eqnarray}

Let us introduce the following operators
\begin{align}
&l_0=\partial^2
&&l_1=a^a\partial_{a\dot{b}}\bar{a}^{\dot{b}}
&&l_1^+=-c^b\partial_{b\dot{a}}\bar{c}^{\dot{a}}.
\label{lll}
\end{align}
Here $\partial_{a\dot{b}}=\sigma^{\mu}_{a\dot{b}}\partial_{\mu}.$
One can show that these operators satisfy the commutation relation
\begin{equation}
[l_1^+,l_1]=(N+\bar{N}+2)l_0,
\label{l1+l1}
\end{equation}
where
\begin{eqnarray}
N=c^aa_a
&\qquad&
\bar{N}=\bar{c}_{\dot{a}}\bar{a}^{\dot{a}}
\end{eqnarray}
All other commutators among these operators (\ref{lll}) vanish.

One can show that in order for a state $|\varphi\rangle$ describe the  continuous spin representation
it is necessary that the following constraints on the vector $|\varphi\rangle$
will satisfied
\begin{eqnarray}
l_0|\varphi\rangle=0
&\qquad&
(l_1-\mu)|\varphi\rangle=0
\label{l0l1}
\end{eqnarray}
where $k=l$ in (\ref{GFState}) is assumed.

Now we turn to construction of the BRST charge and the Lagrangian. Taking
into account that the Lagrangian is real, we should get the
Hermitian BRST charge. However, the system of constraints
(\ref{l0l1}) is not invariant under the Hermitian conjugation. This
situation is similar with BRST Lagrangian construction for free
higher spin fields. Construction of the BRST charge for such a case
was studied in works \cite{Buchbinder1}, \cite{Buchbinder:2005ua}
and we will follow these works. First of all we introduce the
operator $l_{1}^+{-}\mu$ and then add it to the set of constraints
(\ref{l0l1}). Thus set of operators $l_0$, $l_1{-}\mu$,
$l_1^+{-}\mu$ will be invariant under Hermitian conjugation.
Moreover this set of operators will form an algebra with the only
nonzero commutator (\ref{l1+l1})
\begin{eqnarray}
[l_1^+-\mu,l_1-\mu]
=(N+\bar{N}+2)l_0.
\end{eqnarray}
Now we can apply the procedure described in the works
\cite{Buchbinder1}, \cite{Buchbinder:2005ua} and construct Hermitian
BRST charge on the base of operators $l_0$, $l_1{-}\mu$,
$l_1^+{-}\mu$. As a result we arrive at the Hermitian BRST charge in
the form
\begin{eqnarray}
&&
Q=\eta_0l_0+\eta_1^+(l_1-\mu)+\eta_1(l_1^+-\mu)+\eta_1^+\eta_1(N+\bar{N}+2)\mathcal{P}_0
\label{Q}.
\end{eqnarray}
Here we have extended the Fock space by introducing
$\eta_0$, $\eta_1$, $\eta_1^+$ which are the fermionic ghost ``coordinates'' and
$\mathcal{P}_0$, $\mathcal{P}_1^+$, $\mathcal{P}_1$ which are their canonically conjugated ghost ``momenta'' respectively.
These operators obey the anticommutation relations
\begin{eqnarray}
&&
\{\eta_1,{\cal{}P}_1^+\}
= \{{\cal{}P}_1, \eta_1^+\}
=\{\eta_0,{\cal{}P}_0\}
=1
\label{ghosts}
\end{eqnarray}
and act on the vacuum state as follows
\begin{eqnarray}
&&
\eta_1|0\rangle=\mathcal{P}_1|0\rangle=\mathcal{P}_0|0\rangle=0.
\end{eqnarray}
They possess the standard  ghost numbers,
$gh(\mathcal{\eta}^i)$ = $ - gh(\mathcal{P}_i)$ = $1$,
providing the property  $gh(\tilde{Q})$ = $1$.

The operator (\ref{Q}) acts in the extended Fock space of the vectors
\begin{eqnarray}
&&
|\Phi\rangle=
|\varphi\rangle+\eta_0\mathcal{P}_1^+|\varphi_1\rangle+\eta_1^+\mathcal{P}_1^+|\varphi_2\rangle
\label{extened vector}
\end{eqnarray}
and realizes the gauge transformations
\begin{eqnarray}
|\Phi'\rangle = |\Phi\rangle +Q|\Lambda\rangle,
\label{gauge transf}
\end{eqnarray}
for the equation of motion
\begin{eqnarray}
Q|\Phi\rangle=0,
\end{eqnarray}
where $|\Lambda\rangle$ is the gauge parameter
\begin{eqnarray}
|\Lambda\rangle=\mathcal{P}_1^+|\lambda\rangle.
\label{gauge parameter}
\end{eqnarray}
The fields $|\varphi_1\rangle$, $|\varphi_2\rangle$ and and the gauge parameter  $|\lambda\rangle$
in relations (\ref{extened vector}),
(\ref{gauge parameter}) have similar decomposition like
$|\varphi\rangle$ (\ref{GFState}) with $k=l$. In case of $\mu=0$ the
BRST charge (\ref{Q}) becomes BRST charge for the massless higher
spin fields \cite{Buchbinder:2015kca}.

The equations of motion $Q|\Phi\rangle=0$
and gauge transformations $\delta|\Phi\rangle=Q|\Lambda\rangle$ in terms of states $|\varphi_i\rangle$ and
gauge parameter $|\lambda\rangle$ look like
\begin{eqnarray}
&&
l_0|\varphi\rangle-l_1^+|\varphi_1\rangle+\mu|\varphi_1\rangle=0
\\
&&
l_1|\varphi\rangle-l_1^+|\varphi_2\rangle+(N+\bar{N}+2)|\varphi_1\rangle-\mu|\varphi\rangle+\mu|\varphi_2\rangle=0
\label{eq_phi1}
\\
&&
l_0|\varphi_2\rangle-l_1|\varphi_1\rangle+\mu|\varphi_1\rangle=0
\\
&&
\delta|\varphi\rangle=l_1^+|\lambda\rangle-\mu|\lambda\rangle
\qquad
\delta|\varphi_1\rangle=l_0|\lambda\rangle
\qquad
\delta|\varphi_2\rangle=l_1|\lambda\rangle-\mu|\lambda\rangle
\end{eqnarray}

The Lagrangian for the continuous spin field is constructed in the framework of the BRST approach as follows
(see e.g. \cite{Buchbinder:2005ua})
\begin{eqnarray}
{\cal L}
&=&
\int d\eta_0\; \langle\Phi|Q|\Phi\rangle
\;=
\nonumber
\\
&=&
\langle\bar{\varphi}|\Bigl\{l_0|\varphi\rangle-l_1^+|\varphi_1\rangle\Bigr\}
-\langle\bar{\varphi}_1|\Bigl\{l_1|\varphi\rangle-l_1^+|\varphi_2\rangle+(N+\bar{N}+2)|\varphi_1\rangle\Bigr\}
\nonumber
\\
&&{}
-\langle\bar{\varphi}_2|\Bigl\{l_0|\varphi_2\rangle-l_1|\varphi_1\rangle\Bigr\}
\nonumber
\\
&&{}
+\mu\Bigl\{
\langle\bar{\varphi}|\varphi_1\rangle+\langle\bar{\varphi}_1|\varphi\rangle
-\langle\bar{\varphi}_1|\varphi_2\rangle-\langle\bar{\varphi}_2|\varphi_1\rangle
\Bigr\}
\label{action-b}
\end{eqnarray}
Lagrangian (\ref{action-b}) consists of the sum of Lagrangians for
massless bosonic fields plus $\mu$-dependent terms responsible for
the continuous spin.

Now we rewrite the Lagrangian (\ref{action-b}) in terms of the
component fields. Using the equation of motion (\ref{eq_phi1}) we remove the field $|\varphi_1\rangle$
from the Lagrangian (\ref{action-b}). Then calculating the ``average values'' over Fock space vectors, converting the
spin-tensor fields into traceless tensor fields and converting the spin-tensor gauge parameters into traceless tensor field
parameters we arrives at the Lagrangian
\begin{eqnarray}
{\cal L}
&=&
\sum_{s=0}^\infty 2^s \varphi^{\mu(s)}\Bigl[
\partial^2\varphi_{\mu(s)}
-s\,\partial_\mu\partial^\nu\varphi_{\nu\mu(s-1)}
-\frac{s-1}{2}\partial_\mu\partial_\mu\varphi_{2\mu(s-2)}
\nonumber
\\
&&\hspace{10ex}{}
+\frac{\mu^2}{2(s+1)}(\varphi_{\mu(s)}-\varphi_{2\mu(s)})
+\mu\partial^\nu\varphi_{\nu\mu(s)}
\nonumber
\\
&&\hspace{15ex}{}
-\frac{\mu}{2}\partial_\mu\varphi_{\mu(s-1)}
+\frac{\mu}{2}\;\frac{2s+1}{s+1}\partial_\mu\varphi_{2\mu(s-1)}
\Bigr]
\nonumber
\\
&&{}
-\sum_{s=0}^\infty 2^s \varphi^{\mu(s)}_2\Bigl[
\frac{2s+3}{s+2}\partial^2\varphi_{2\,\mu(s)}
+\frac{s^2}{s+2}\partial_\mu\partial^\nu\varphi_{2\nu\mu(s-1)}
+2(s+1)\partial^\nu\partial^\tau\varphi_{\nu\tau\mu(s)}
\nonumber
\\
&&\hspace{10ex}{}
+\frac{\mu^2}{2(s+1)}(\varphi_{\mu(s)}-\varphi_{2\mu(s)})
+\mu\frac{2s+3}{s+2}\partial^\nu\varphi_{\nu\mu(s)}
\nonumber
\\
&&\hspace{10ex}{}
-\mu\frac{s+1}{s+2}\partial^\nu\varphi_{2\nu\mu(s)}
+\frac{\mu}{2}\;\frac{s}{s+1}\partial_\mu\varphi_{2\mu(s-1)}
\Bigr]
\label{action-b2}
\end{eqnarray}
and gauge transformations
\begin{eqnarray}
&&
\delta\varphi_{\mu(s)}=s\,\partial_{\mu_s}\lambda_{\mu(s-1)}
-\frac{s-1}{2}\,\eta_{\mu(2)}\,\partial^\nu\lambda_{\nu\mu(s-2)}
-\mu\,\lambda_{\mu(s)},
\label{gauge1}
\\[0.5em]
&&
\delta\varphi_{2\,\mu(s)}=-2(s+1)\,\partial^{\nu}\lambda_{\nu\mu(s)}-\mu\,\lambda_{\mu(s)}.
\label{gauge2}
\end{eqnarray}
We note that the set of the fields in Lagrangian (\ref{action-b2}) is the same like in the Metsaev's Lagrangian
(2.14) in \cite{Metsaev:2018lth} in the case $d=4$, $m=0$. These Lagrangians will be exactly the same if we make the
redefinition of the fields $\varphi^{\mu(n)}\to{}A_n\phi_I^{a(n)}$,  $\varphi_2^{\mu(n)}\to{}-A_n\phi_{II}^{a(n)}$
and gauge parameters $\lambda^{\mu(n)}\to{}A_{n+1}\xi^{a(n)}$
where $A_n=(2^{n+1}n!)^{-1/2}$ and also $\mu\to\kappa$. As a result we conclude that the BRST Lagrangian construction works
perfectly for the continuous spin fields as well as for all other higher spin fields.

\section{Summary}
We have developed the Lagrangian BRST construction for the continuous spin field theory.
\begin{itemize}
\item{We have reformulated the Wigner equations, defining the irreducible representation with continuous spin,
in terms of two auxiliary bosonic 2-component spinor variables. The spin operator (\ref{M}) for such an representation has
been introduced. The representation is realized on fields satisfying the constraints
(\ref{C1}), (\ref{C2}), (\ref{C3}). The number of the constraints
turns out to be less than in case of the usually used representation in terms of vectorial auxiliary variable. }
\item{The constraints are reformulated in terms of operators acting on the vectors of the auxiliary Fock space.
Extra operator has been introduced to provide the real Lagrangian.
The algebra of all operastors has been calculated.}
\item{Hermitian BRST charge is constructed (\ref{Q}) for the continuous spin field theory, the Lagrangian and gauge
transformations in terms of the Fock space vectors (\ref{action-b}), and in terms of traceless tensor fields
(\ref{action-b2}), (\ref{gauge1}), (\ref{gauge2}) have been derived. The Lagrangian
coincides with Metsaev Lagrangian \cite{Metsaev:2018lth} after some redefinitions of the fields and
gauge parameters.}

\end{itemize}
Thus, the BRST Lagrangian construction, developed in refs.
\cite{Buchbinder1,Buchbinder:2004gp,Buchbinder:2005ua,Buchbinder:2006nu,Buchbinder:2006ge}
is generalized for the continuous spin field.

The results of the paper can be applied for deriving the Lagrangians for fermionic continuous spin field and for
supersymmetric continuous spin field theory. It would be also interesting to generalize these results for the continuous
spin fields in the AdS space.

\section*{Acknowledgments}

The authors are thankful to R.R. Metsaev for useful comments. I.L.B is appreciative to A.P. Isaev and S. Fedoruk for
discussing the various aspects of the continuous spin fields. The research was supported in parts by Russian Ministry
of Education and Science, project No.~3.1386.2017. The authors are also grateful to RFBR grant,
project No.~18-02-00153 for partial support.

\end{document}